\documentclass[reprint,nofootinbib,amsmath,amssymb,aps,floatfix,superscriptaddress,twocolumn]{revtex4-2}
\usepackage{graphicx}
\usepackage{dcolumn}
\usepackage[colorlinks,linkcolor=magenta,anchorcolor=cyan,citecolor=blue]{hyperref}
\usepackage{bm}
\usepackage[multiple]{footmisc}

\begin{document}
\title{Energy extraction from rotating regular black hole via Comisso-Asenjo mechanism}
\author{Zhen Li} 
\email{zhen.li@nbi.ku.dk}
\affiliation{DARK, Niels Bohr Institute, University of Copenhagen, Jagtvej 128, 2200 Copenhagen Ø, Denmark}
\author{Xiao-Kan Guo}
\email{kankuohsiao@whu.edu.cn }
\affiliation{Department of Applied Mathematics, Yancheng Institute of Technology, Yancheng 224051, China}
\author{Faqiang Yuan}
\email{yfq@mail.bnu.edu.cn}
\affiliation{Department of Physics, Beijing Normal University, Beijing 100875, China}

\date{\today}
\begin{abstract}
Recently, it has been demonstrated by Comisso and Asenjo that magnetic reconnection processes in the ergosphere of a Kerr black hole can provide us with a promising mechanism for extracting the rotational energy from it. In this paper, we study the energy extraction from the the newly proposed rotating regular black holes via this Comisso-Asenjo mechanism. This novel rotating regular black hole has an exponential convergence factor $e^{-k/r}$ on the mass term characterized by the regular parameter $k$ in the exponent. We explore the effects of this regular parameter on the magnetic reconnection as well as other critical parameters determining the Comisso-Asenjo process. The parameter spaces allowing energy extraction to occur are investigated. The power, efficiency and the power ratio to the Blandford-Znajek mechanism are studied. The results show that the regularity of the rotating black hole has significant effects on the energy extraction via the Comisso-Asenjo mechanism.
\end{abstract}
\maketitle

\section{Introduction}
General relativity is the theory that best captures our present understandings of gravitational interaction. Gravitational waves \cite{gw1,gw2,gw3} and black hole images \cite{shadow1, shadow2} have recently been observed, adding to the compelling evidences for general relativity. However, general relativity also encounters a number of difficulties and issues, of which the singularity problem  in classical general relativity \cite{cha2,cha3} is the most severe one. It is generally accepted that singularities do not exist in nature and instead show the limits of general relativity. The regular black holes, which are solutions with horizons but are non-singular in the sense that curvature invariants are finite everywhere, can offer an alternative solution for the singularity problem. In addition to the typical regular black hole solutions such as the Bardeen and Hayward regular black holes \cite{re,re2,Hay06}, many new regular black holes are derived and studied recently (cf. the review \cite{LYGM23}).
To construct such regular black hole solutions, one could modify the shape function  in the black hole metric in such a way that the the curvature invariants are finite.
A highly efficient way of obtaining regular solution is to multiply the mass function by an exponential factor $e^{-k/r}$, and many new regular black hole metrics have been generated by this approach \cite{nre,nre1,nre2,Cul17,Cul22,Cul23,nre5,nre4,nre6,nre7}. In particular, the regular rotating black hole obtained this way \cite{nre5,nre6,nre7}, which are different from those obtained by the Newman-Janis algorithm \cite{re4}, have attracted  lots of attention, 
but the phenomenological studies of such black holes are still very lacking. There are a few works exploring this aspect, mainly focusing on the black hole image and quasinormal mode effects \cite{fw1,fw2,fw3,fw4,SAAA22}, and therefore other astrophysical tests of regular rotating black holes are desirable.

A rapidly rotating black hole will produce an antiparallel magnetic field configuration in the equatorial plane \cite{mr1,mr2}. Both numerical simulations \cite{mr3,mr4,mr5,mr6} and black hole image \cite{shadow1,shadow2} support that the main condition for this configuration is realistic. Comisso and Asenjo has shown in their remarkable paper \cite{mr11} that this peculiar magnetic field configuration could lead to a fast magnetic reconnection process inside the ergosphere when the aspect ratio of the current sheet exceeds critical value \cite{mr7,mr8,mr9}, which can convert an amount of magnetic energy into plasma particle energy so that the plasma can escape from the reconnection layer.  We will denote this new mechanism as the Comisso-Asenjo mechanism in order to distinguish with previous attempts in magentic reconnection \cite{mr10}. Comisso-Asenjo mechanism offers us a brand new way to extract energy from the rotating black holes. It has also been shown in many numerical simulations that there is always a dominant point at which the reconnection process occurs \cite{mr3,mr5,mr6}. In this Comisso-Asenjo magnetic reconnection process, one part of the plasma or flux is accelerated and another part is decelerated in the opposite direction. If the decelerated part has a negative energy and the accelerated part has an energy greater than its rest mass and thermal energy at infinity, the energy is extracted from the rotating black hole by magnetic reconnection \cite{mr11}. The frame dragging effect of a rapidly rotating black hole causes this process to occur over and over again. In comparison with other energy extraction mechanisms, such as the Penrose process  \cite{p1,p2} and the Blandford-Znajek mechanism \cite{bz4}, such a mechanism could be dominant in extracting the rotational energy of black holes.

The energy extraction via Comisso-Asenjo mechanism was first studied in the Kerr black hole \cite{mr11} and recently extended to many other rotating  black holes \cite{mrr1,mrr2,mrr3,mrr4,mrr5}. All the results show that the effects on the reconnection process of non-Kerr black hole are significantly different to the Kerr case. In this work, we aim to investigate the energy extraction from the rotating regular black hole via Comisso-Asenjo mechanism in the ergoshpere. Because this process could potentially produce more high-energy astrophysical phenomena, it allows us to test the rotating regular black hole hypothesis across a wider observational range.

This paper is organized as follows: We introduce the rotating regular black hole spacetime in Sec.\ref{sec2}. Then in Sec.\ref{sec3}, we will present the formulations of Comisso-Asenjo mechanism, especially the equations of plasma energy-at-infinity density per enthalpy and the conditions for energy extraction from a regular rotating black hole to occur. Next, based on the formulas in the last section, in Sec.\ref{sec4}, we will explore the parameter spaces allowing energy extraction from the rotating regular black hole via Comisso-Asenjo mechanism. In Sec.\ref{sec5}, we study the power and efficiency of this mechanism with different parameter combinations. We also compare the power ratio between Comisso-Asenjo mechanism and the Blandford-Znajek mechanism in this section. 
In Section.\ref{sec6}, we will make a conclusion.

\section{Rotating regular black hole}\label{sec2}
The metric of rotating regular black hole that we will discuss can be written in the Boyer-Lindquist coordinates as
\cite{nre6,nre7}, 
\begin{equation}\label{metric}
d s^{2}=g_{t t} d t^{2}+g_{r r} d r^{2}+g_{\theta \theta} d \theta^{2}+g_{\phi \phi} d \phi^{2}+2 g_{t \phi} d t d \phi
\end{equation}
where the metric components are given by 
\begin{align}
g_{t t}&= -\left(1-\frac{2 M r \mathrm{e}^{-k / r}}{\Sigma}\right)\quad   g_{t\phi}=-\frac{2 a M r \mathrm{e}^{-k / r}}{\Sigma} \sin ^{2} \theta \nonumber\\
g_{r r} &=\frac{\Sigma}{\Delta}\qquad \qquad \quad\quad\quad\quad\quad g_{\theta \theta} =\Sigma \nonumber\\
g_{\phi\phi}&=\left(r^{2}+a^{2}+\frac{2 M r a^{2} \mathrm{e}^{-k / r}}{\Sigma} \sin ^{2} \theta\right) \sin ^{2} \theta 
\end{align}
with $\Sigma=r^{2}+a^{2} \cos ^{2} \theta$ and $\Delta=r^{2}+a^{2}-2 M re^{-k / r}$. The mass, specific angular momentum, and regular parameters,  $M$, $a$, and $k$ are assumed to be positive. The Kerr metric could be retained when we set $k/r=0$. Here and after, we use geometrized units with $G=c =1$.

This spacetime is regular everywhere. We can see this from the curvature invariants, for example, the Kretschmann invariant ${K}=R_{a b c d} R^{a b c d}\left(R_{a b c d}\right.$ is the Riemann tensor):
\begin{equation}
K=\frac{4 M^{2} \mathrm{e}^{\frac{-2 k}{r}}}{r^{6} \Sigma^{6}}\left(\Sigma^{4} k^{4}-8 r^{3} \Sigma^{3} k^{3}+A k^{2}+B k+C\right)
\end{equation}
where $A$, $B$, and $C$ are functions of $r$ and $\theta$,
\begin{equation}
\begin{array}{l}
A=-24 r^{4} \Sigma\left(-r^{4}+a^{4} \cos ^{4} \theta\right) \\
B=-24 r^{5}\left(r^{6}+a^{6} \cos ^{6} \theta-5 r^{2} a^{2} \cos ^{2} \theta \Sigma\right) \\
C=12 r^{6}\left(r^{6}-a^{6} \cos ^{6} \theta\right)\\
\quad\quad-180 r^{8} a^{2} \cos ^{2} \theta\left(r^{2}-a^{2} \cos ^{2} \theta\right).
\end{array}
\end{equation}
For $M \neq 0$, they are regular everywhere.

The solutions of equation 
\begin{equation}\label{eh}
\Delta=r^{2}+a^{2}-2 M re^{-k / r}=0
\end{equation}
give  the horizons.  However, there are no analytical solutions. The numerical results of horizon structure as well as the ergosphere with different parameters were discussed in \cite{nre5}. We will restrict our discussion on the value of regular parameter $k$ such that (\ref{eh}) has two distinct real solutions (outer/event horizon and inner horizon), i.e, $k$ should be less than the critical value $k_c^{EH}$ which decreases as $a$ increases.

\section{Energy extraction via Comisso-Asenjo mechanism}\label{sec3}
In this section, we present the Comisso-Asenjo formulas \cite{mr11} of calculating the energy at infinity associated with accelerated/decelerated plasma in the spacetime \eqref{metric}. It is more convenient to evaluate the plasma energy density in the  “zero-angular-momentum-observer” (ZAMO) frame \cite{zamo}. The metric (\ref{metric}) in ZAMO frame takes the form of a Minkowski metric
\begin{equation}
d s^{2}=-d \hat{t}^{2}+\sum_{i=1}^{3}\left(d \hat{x}^{i}\right)^{2}=\eta_{\mu \nu} d \hat{x}^{\mu} d \hat{x}^{\nu}
\end{equation}
where
\begin{equation}
d \hat{t}=\alpha d t, \quad d \hat{x}^{i}=\sqrt{g_{i i}} d x^{i}-\alpha \beta^{i} d t
\end{equation}
with
\begin{equation}
\alpha=\left(-g_{t t}+\frac{g_{\phi t}^{2}}{g_{\phi \phi}}\right)^{1 / 2},\quad \beta^{\phi}=\frac{\sqrt{g_{\phi \phi}} \omega^{\phi}}{\alpha}.
\end{equation}
We define $\omega^{\phi}=-g_{\phi t} / g_{\phi \phi}$ as the angular velocity of the frame dragging due to the rotating regular spacetime.

For a contravariant vector $a^{\mu}$ in the Boyer-Lindquist coordinates, when transformed into the ZAMO frame, which we denote by $\hat{a}^{\mu}$, the following relation is obtained
\begin{equation}
\hat{a}^{0}=\alpha a^{0}, \quad \hat{a}^{i}=h_{i} a^{i}-\alpha \beta^{i} a^{0}
\end{equation}
and we also have the covariant vector $\hat{a}_{\mu}$ transformation relations
\begin{equation}
\hat{a}_{0}=\frac{1}{\alpha} a_{0}+\sum_{i} \frac{\beta^{i}}{h_{i}} a_{i}, \quad \hat{a}_{i}=\frac{1}{h_{i}} a_{i}.
\end{equation}

The one-fluid approximation energy-momentum tensor of this system, in Boyer-Lindquist coordinates, takes form of
\begin{equation}
T^{\mu \nu}=pg^{\mu \nu}+w U^{\mu}U^{\nu}+F^{\mu}_{\,\,\,\delta}F^{\nu \delta}-\frac{1}{4}g^{\mu \nu}F^{\rho \delta}F_{\rho \delta}
\end{equation}
where $p$, $w$, $U^{\mu}$, and $F_{\mu \nu}$ are respectively the proper plasma pressure, enthalpy density, four-velocity, and electromagnetic field tensor. With this energy-momentum tensor, we can get the  “energy-at-infinity” density $e^{\infty}=-\alpha g_{\mu 0}T^{\mu 0}=e^{\infty}_{hyd}+e^{\infty}_{em}$ \cite{mr10}, where $e^{\infty}_{hyd}$ and $e^{\infty}_{em}$ are respectively the hydrodynamic energy-at-infinity density and the electromagnetic energy-at-infinity density, and they are given by
\begin{align}
e^{\infty}_{hyd}&=\alpha(w \hat \gamma^2-p)+\alpha\beta^{\phi}w\hat \gamma^2 \hat v^{\phi}\nonumber\\
e^{\infty}_{em}&=\frac{\alpha}{2} (\hat B^2+\hat E^2)+(\hat{\bf B}\times \hat{\bf E})_{\phi}
\end{align}
where $\hat \gamma=\hat U^0=1/\sqrt{1-{\textstyle \sum_{i=1}^{3}} (d \hat v^i)^2}$, $\hat B^i=\epsilon ^{ijk}\hat F_{jk}/2$, $\hat E^i=\hat F_{i0}$ are the Lorentz factor, the components of magnetic and electric fields respectively. Here, $v^{\phi}$ denotes the azimuthal component of the plasma outflow velocity in the ZAMO frame \cite{AC17}. The hat above these quantities means that we evaluate the quantity in the ZAMO frame. 

Considering that a considerable portion of the magnetic energy is converted into kinetic energy of plasma during the Comisso-Asenjo magnetic reconnection process, one can ignore the contribution of $e^{\infty}_{em}$ in the total energy, which leads to $e^{\infty}\approx e^{\infty}_{hyd}$. In addition, considering the plasma element is incompressible and adiabatic, we have \cite{mr10}
\begin{equation}\label{a1}
e^{\infty}= \alpha\left(w(\hat \gamma +\beta^{\phi}\hat \gamma\hat v^{\phi})+\frac{p}{\hat \gamma}\right).
\end{equation}

We would like to introduce the local rest frame in order to investigate the localized magnetic reconnection process, $\bar x^{\mu}=(\bar x^{0}, \bar x^{1}, \bar x^{2}, \bar x^{3})$, in which the directions of $ \bar x^{1}$ and $ \bar x^{3}$ are repectively parallel to the radial direction $\bar x^{1}= r$ the azimuthal direction $ \bar x^{3}=\phi$. The local rest frame of plasma rotates with Keplerian angular velocity $\Omega_K$ in the equatorial plane from the perspective of Boyer-Lindquist observer, which is given by
\begin{equation}
\Omega_K=\frac{d \phi}{d t}=\frac{-g_{t \phi, r}+\sqrt{g_{t \phi, r}^{2}-g_{t t, r} g_{\phi \phi, r}}}{g_{\phi \phi, r}}.
\end{equation}
Based on the transformation relation of vectors between Boyer-Lindquist and ZAMO coordinates, we can also obtain the above Keplerian angular velocity observed in the ZAMO frame, which is given by
\begin{equation}
\hat v^K= \frac{\Omega_K\sqrt{g_{\phi\phi}}}{\alpha}-\beta^{\phi}.
\end{equation}
If we denote the outflow velocity observed in the local rest frame as $v_{out}$, then the outflow velocity observed in the ZAMO frame
is
\begin{equation}\label{a2}
\hat v^{\phi}_{\pm}=\frac{\hat v^K\pm v_{out}cos(\xi) }{1\pm \hat v_K v_{out}cos(\xi)}
\end{equation}
where $\pm$ represent the outflow velocity with corotating (+) and counterrotating (-) direction relative to the rotation of the black hole. They also correspond to the accelerated part and decelerated part of the plasma respectviely. In \eqref{a2}, $\xi=arctan(\bar v^1/\bar v^3)$ is the plasma orientation angle, $\bar v^1$ and $\bar v^3$ are the radial and azimuthal components of plasma velocities in the local rest frame.

With the above equations (\ref{a1}) and (\ref{a2}), we can get the energy-at-infinity density of the reconnection outflows as
\begin{align}
e_{\pm}^{\infty}= & \alpha \hat{\gamma}_{K}\left(\left(1+\hat{v}_{K} \beta^{\phi}\right) \gamma_{{out}} w\right.\nonumber\\
& \pm \cos (\xi)\left(\hat{v}_{K}+\beta^{\phi}\right) \gamma_{{out}} v_{{out}} w \nonumber\\
& \left.-\frac{p}{\left(1 \pm \cos (\xi) \hat{v}_{K} v_{{out}}\right) \gamma_{{out}} \hat{\gamma}_{K}^{2}}\right)
\end{align}
where $\gamma_{{out}}=1/\sqrt{1-v_{out}^2}$ and $\hat{\gamma}_{K}=1/\sqrt{1-\hat{v}_{K}^2}$.
From \cite{mr11}, Comisso and Asenjo have derived that $v_{out}$ is related to the properties of plasma magnetization and it could be expressed as
\begin{equation}
v_{out}=\sqrt{\frac{\sigma_0}{\sigma_0+1}}
\end{equation}
where $\sigma_0= B_0^2/w_0$ is the plasma magnetization upstream of the reconnection layer, $B_0$ is the asymptotic macro-scale magnetic field and $w_0$ is the enthalpy density of the plasma.
Then, the plasma energy-at-infinity density per enthalpy $\epsilon_{\pm}^{\infty}=e_{\pm}^{\infty}/w$ becomes \cite{mr11} 
\begin{align}\label{ep}
\epsilon_{ \pm}^{\infty}=\alpha \hat{\gamma}_{K}[& \left(1+\beta^{\phi} \hat{v}_{K}\right)\left(1+\sigma_{0}\right)^{1 / 2} \pm \cos (\xi)\left(\beta^{\phi}+\hat{v}_{K}\right) \sigma_{0}^{1 / 2} \nonumber\\
& \left.-\frac{1}{4} \frac{\left(1+\sigma_{0}\right)^{1 / 2} \mp \cos (\xi )\hat{v}_{K} \sigma_{0}^{1 / 2}}{\hat{\gamma}_{K}^{2}\left(1+\sigma_{0}-\cos ^{2} (\xi) \hat{v}_{K}^{2} \sigma_{0}\right)}\right]
\end{align}
where we have assumed that $p=w/4$. Eq.(\ref{ep}) has exactly the same form as the one in the Kerr black hole case (see \cite{mr11}), and the differences are the geometry quantities which are now replaced by spacetime metric (\ref{metric}).

Just like in the Penrose process \cite{p1,p2}, if the following conditions should be satisfied
\begin{equation}\label{cd}
\epsilon_{-}^{\infty}<0 ,\quad
\Delta\epsilon_{+}^{\infty}=\epsilon_{+}^{\infty}-\left(1-\frac{\Gamma}{4(\Gamma-1)} \right)>0    
\end{equation}
for a relativistic hot plasma, i.e. $\Gamma=4/3$, we have $\Delta\epsilon_{+}^{\infty}=\epsilon_{+}^{\infty}$. Consequently, black hole energy can only be extracted if the decelerated part of plasma in a magnetic reconnection process acquires negative energy as measured at infinity, while the accelerated part of the plasma in the same magnetic reconnection process acquires energy at infinity larger than its rest mass and thermal energies.

\section{Parameter spaces for energy extraction }\label{sec4}
The expression of energy at infinity Eq.(\ref{ep}) depends on several critical parameters: the black hole mass $M$, the black hole spin $a$, the dominant reconnection radial location $r$ (so-called X point \cite{mr3,mr5,mr6,mr11}), the plasma magnetization $\sigma_0$, the orientation angle $\xi$ and the regular parameter $k$. For simplicity, we will choose units by setting $M = 1$ in the rest of this paper, then $a$, $r$ and $k$ are measured in the unit of $M$, while $\sigma_0$ and $\xi$ are dimensionless parameters.

Now, we will show that Comisso-Asenjo mechanism is a viable mechanism to extract energy from regular rotating black holes in a significant range of parameter spaces.  In order to compare with the literature, the structure of figures in this section are in reference to the Comisso and Asenjo original paper \cite{mr11}.

\begin{figure}[t]
  \includegraphics[scale=0.6]{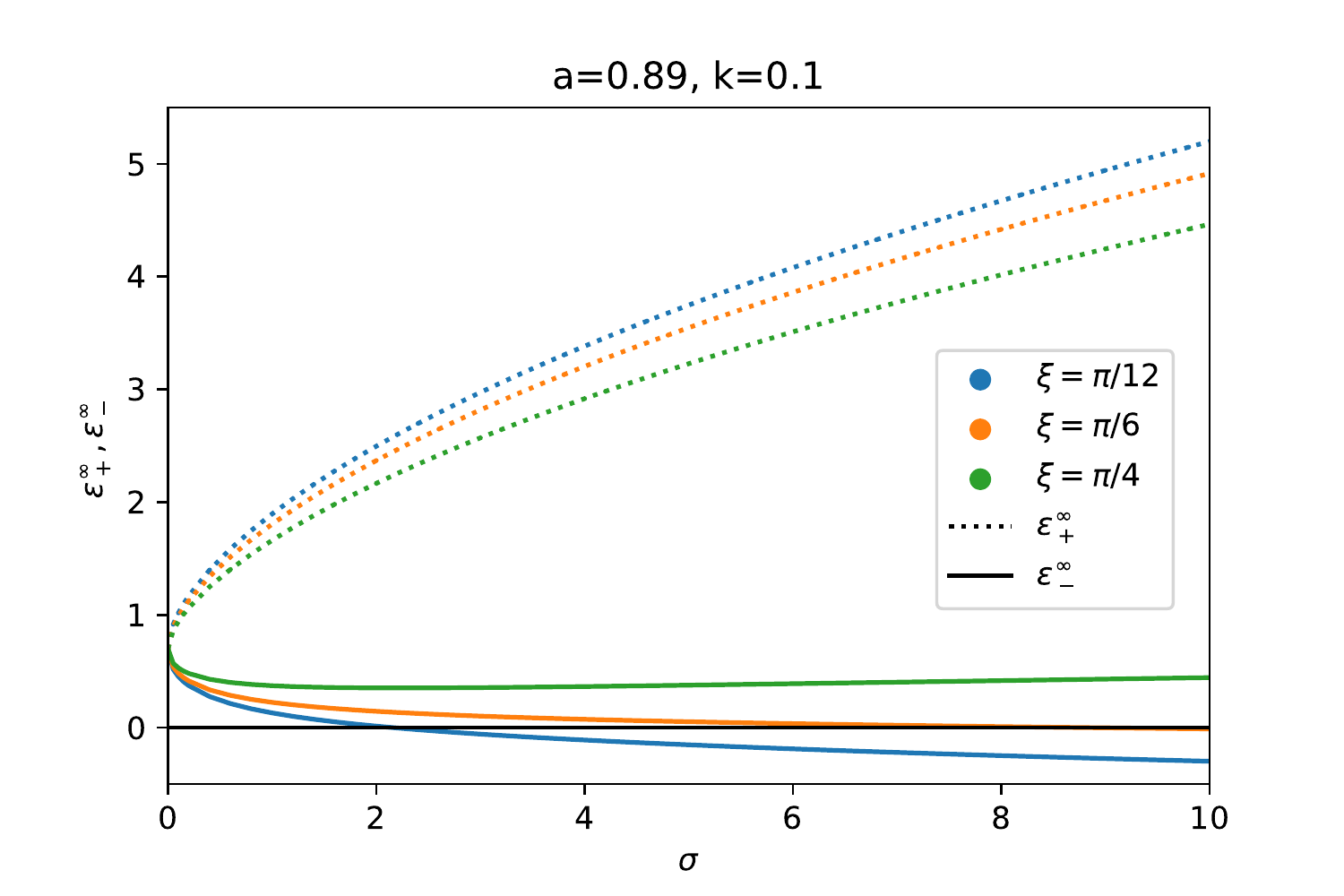}
\caption{The behaviors of $\epsilon_{+}^{\infty}$ (dotted curve) and $\epsilon_{-}^{\infty}$ (solid curve) as a function of plasma magnetization $\sigma_0\in[0,10]$, with different orientation angle $\xi=\pi/12,\pi/6,\pi/4$. The dominant reconnection radial location is taken as $r=1.5$, with black hole spin $a=0.89$ and regular parameter $k=0.1$. The black solid line is $\epsilon_{+}^{\infty}=\epsilon_{-}^{\infty}=0$ as reference.}
\label{es}
\end{figure}

First, let us consider how the orientation angle $\xi$ and plasma magnetization $\sigma_0$ affect the energy-at-infinity per enthalpy $\epsilon_{+}^{\infty}$ and $\epsilon_{-}^{\infty}$. For this purpose, taking $r=1.5$, $a=0.89$ and $k=0.1$, we plot $\epsilon_{+}^{\infty}$ and $\epsilon_{-}^{\infty}$ as a function of $\sigma_0$ with different $\xi$ in Fig.\ref{es}.
We can see that $\epsilon_{+}^{\infty}$ increase with the plasma magnetization $\sigma_0$ while $\epsilon_{-}^{\infty}$ decrease with the plasma magnetization $\sigma_0$. 
It is easy to satisfy the condition $\Delta \epsilon_{+}^{\infty}=\epsilon_{+}^{\infty}>0$, however, the orientation angle is essential for $\epsilon_{-}^{\infty}$ to be negative. In order to satisfy the energy extraction conditions (\ref{cd}), the orientation angle should be small enough. The restriction could be reduced if the plasma magnetization $\sigma_0$ is large enough since it can subtract the increase of $\epsilon_{-}^{\infty}$ due to increase of orientation angle $\xi$.

\begin{figure}[t]
  \includegraphics[scale=0.6]{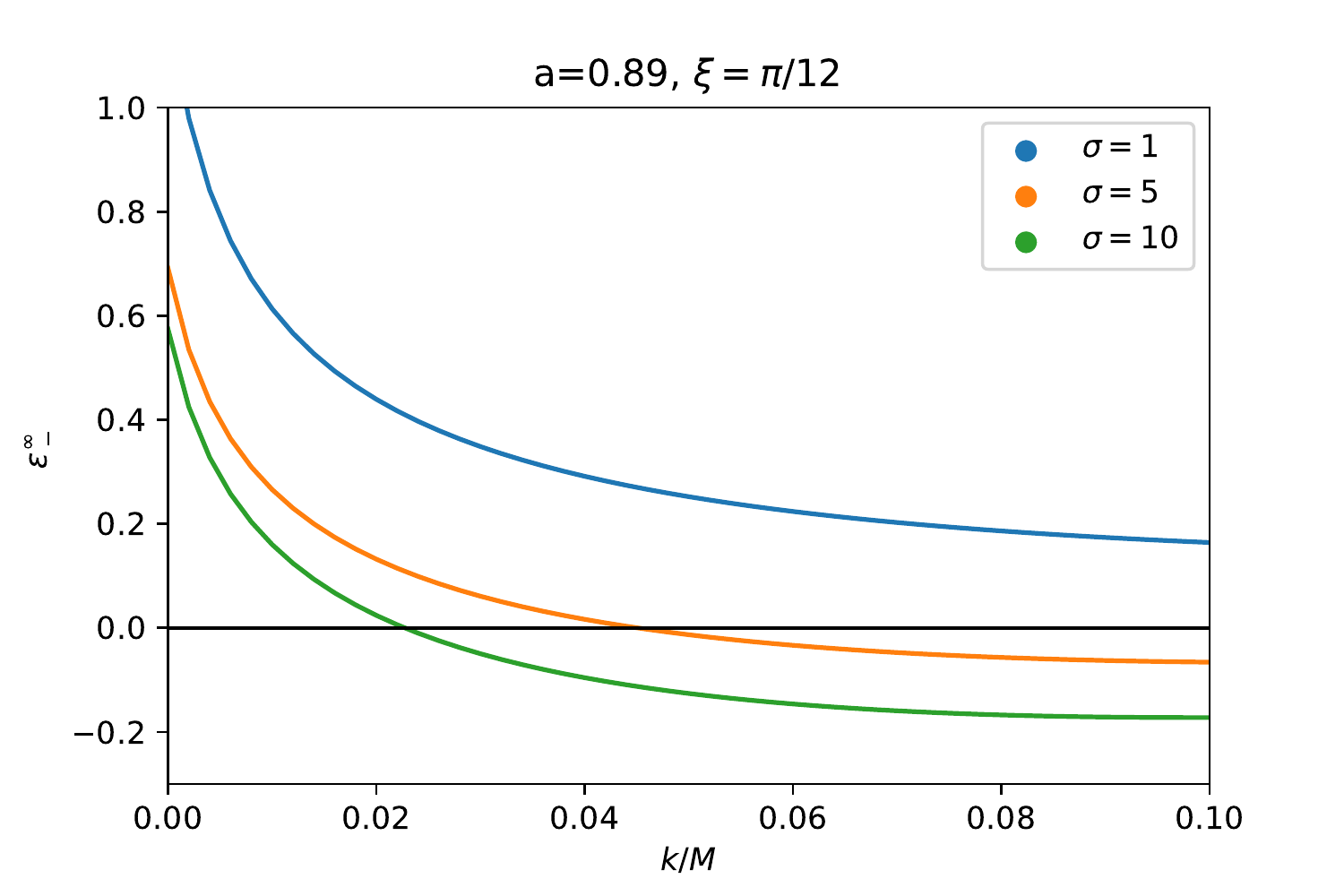}  
  \includegraphics[scale=0.6]{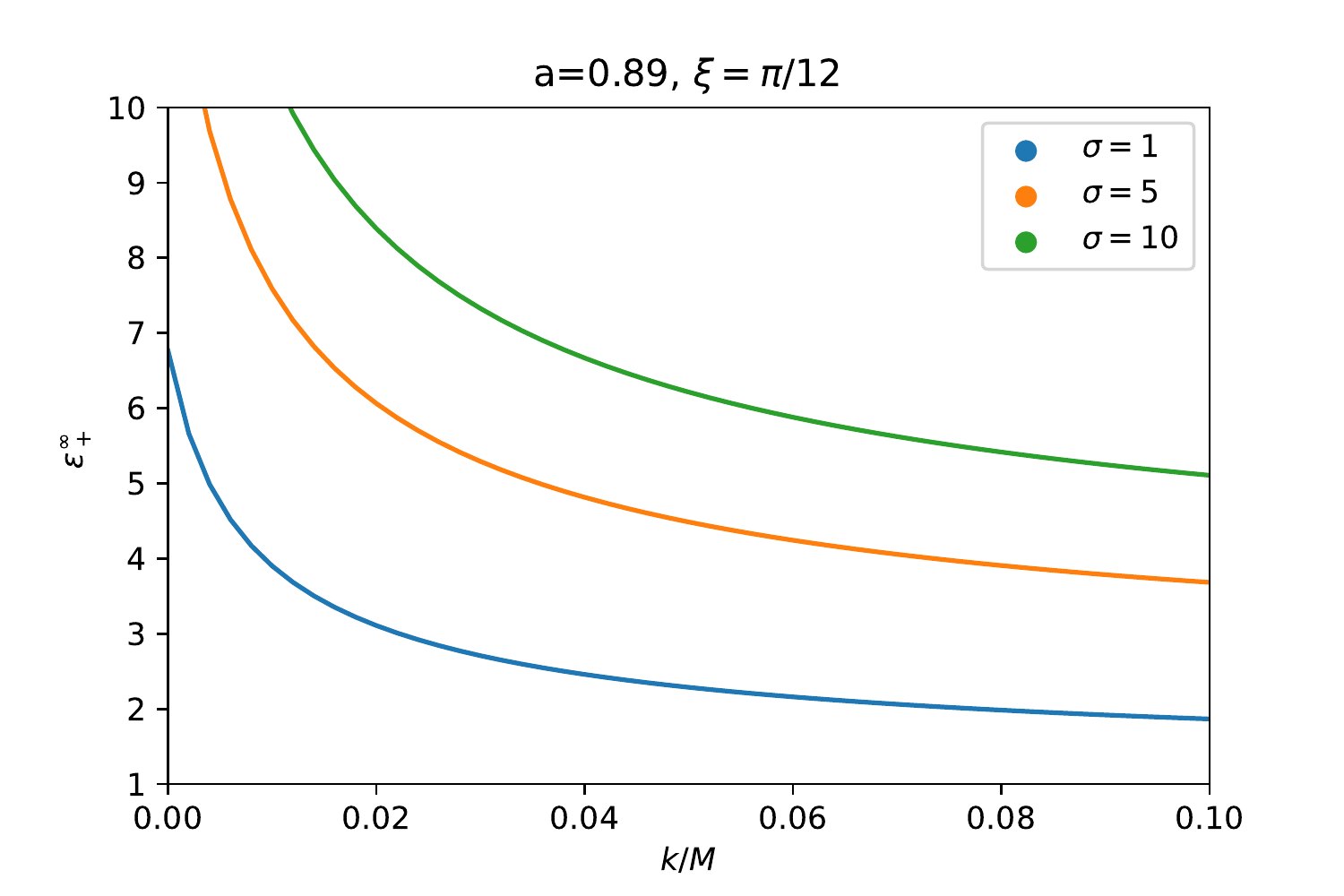}
\caption{The behaviors of $\epsilon_{-}^{\infty}$ (upper plot) and $\epsilon_{+}^{\infty}$ (lower plot) as a function of the regular parameter
$k\in[0,0.1]$, with different plasma magnetization $\sigma_0=1,5,10$. The dominant reconnection radial location is taken as $r=1.6$, with black hole spin $a=0.89$ and orientation angle $\xi=\pi/12$. The black solid line in the upper plot is $\epsilon_{-}^{\infty}=0$ as reference.}
\label{ek}
\end{figure}

To investigate the impact of regular parameter $k$ on $\epsilon_{+}^{\infty}$ and $\epsilon_{-}^{\infty}$, we plot $\epsilon_{+}^{\infty}$ and $\epsilon_{-}^{\infty}$ as a function of regular parameter $k$ in Fig.\ref{ek} by taking $r=1.6$, $a=0.89$ and $\xi=\pi/12$ with different plasma magnetization $\sigma_0$. From both upper plot and lower plot, we can see that $\epsilon_{+}^{\infty}$ and $\epsilon_{-}^{\infty}$ all decrease as the regular parameter $k$ increases. Again, $\epsilon_{+}^{\infty}>0$ is well satisfied regardless what the plasma magnetization $\sigma_0$ is. However, in order to extract energy from the black hole, the condition $\epsilon_{-}^{\infty}<0$ is not so straightforward to be met. It requires that the regular parameter $k$ should be large enough. What's more, from both plots, we can see that a large plasma magnetization $\sigma_0$ is always beneficial for extracting energy from the black hole.

\begin{figure}[htbp]
  \includegraphics[scale=0.6]{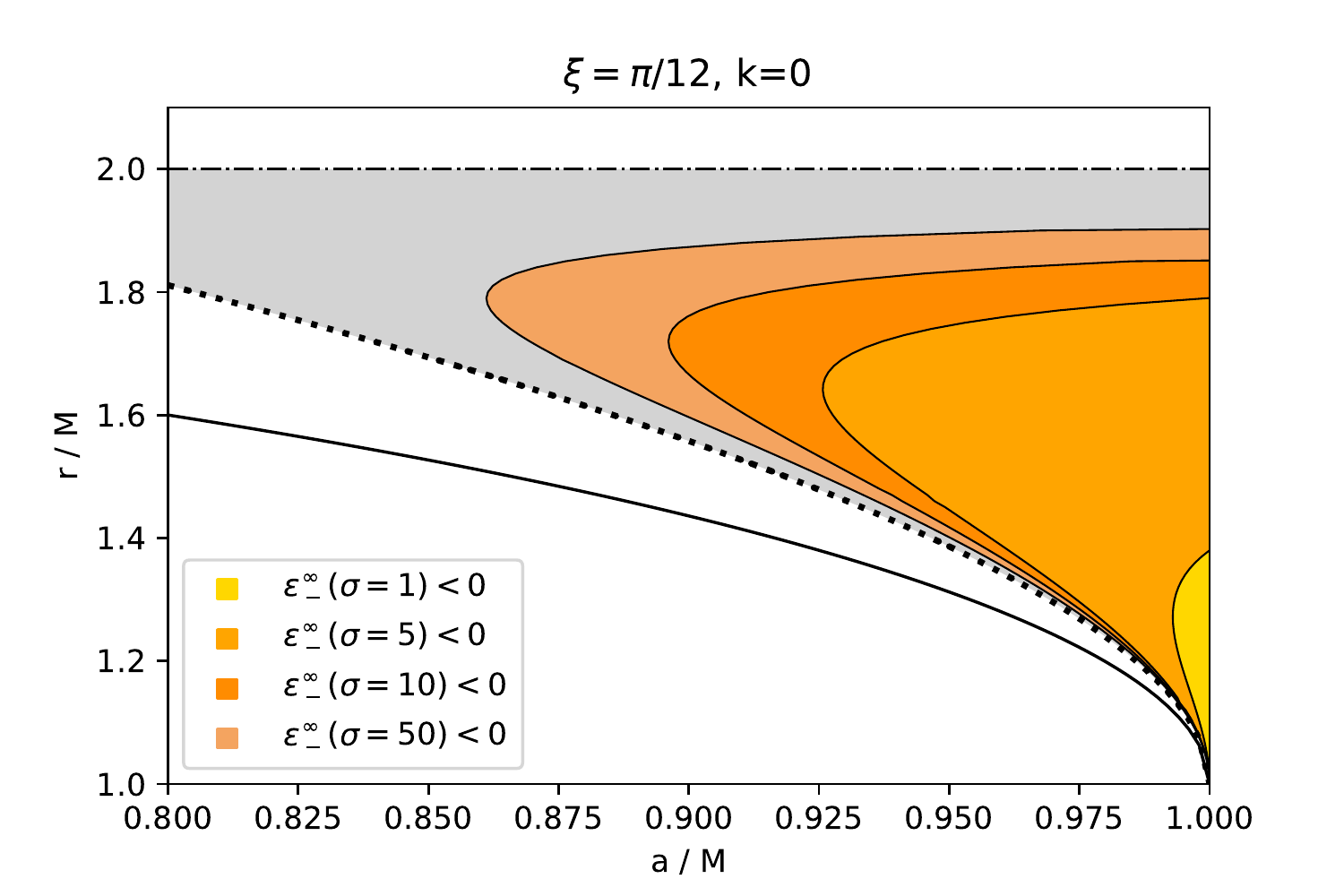}
  \includegraphics[scale=0.6]{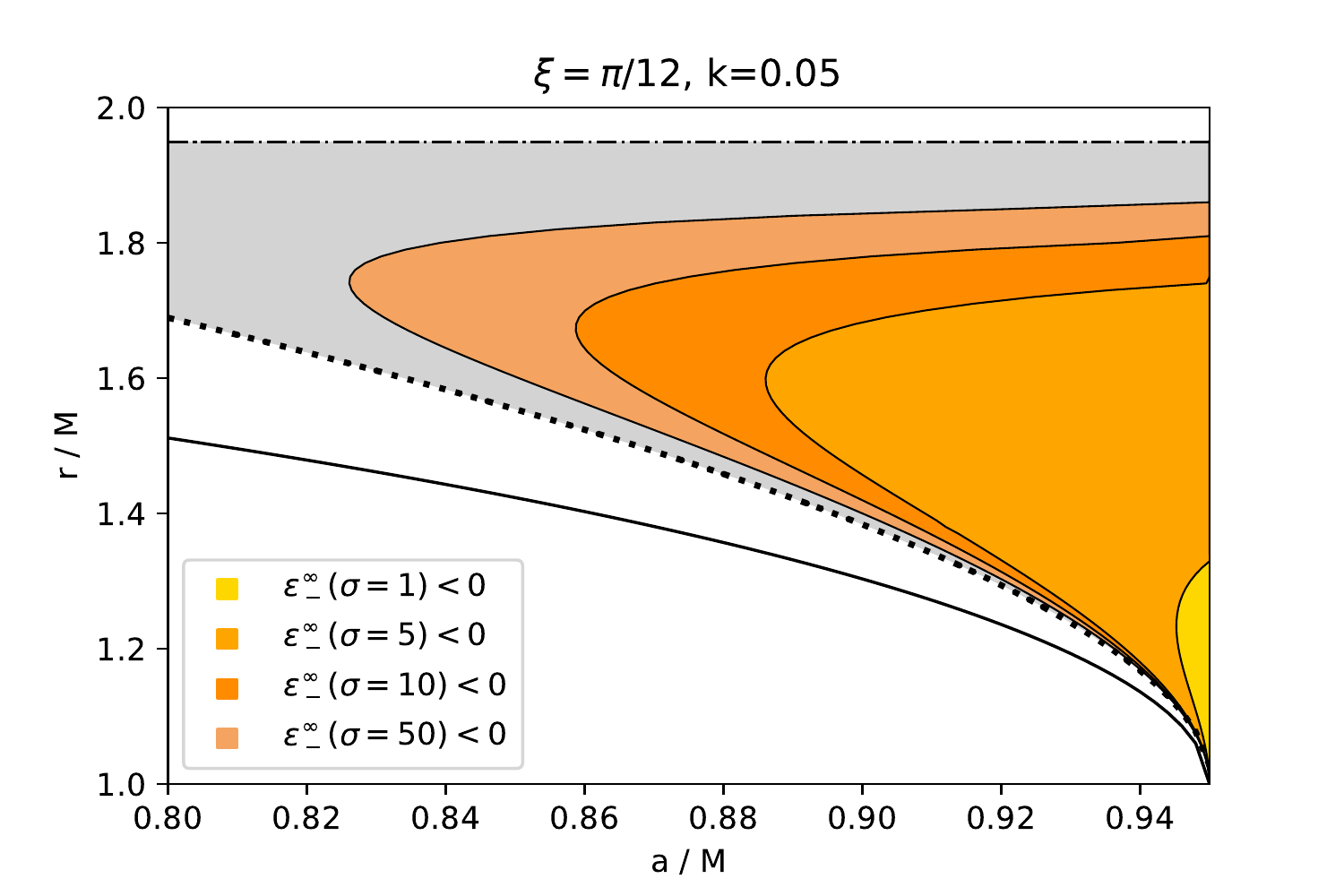}
  \includegraphics[scale=0.6]{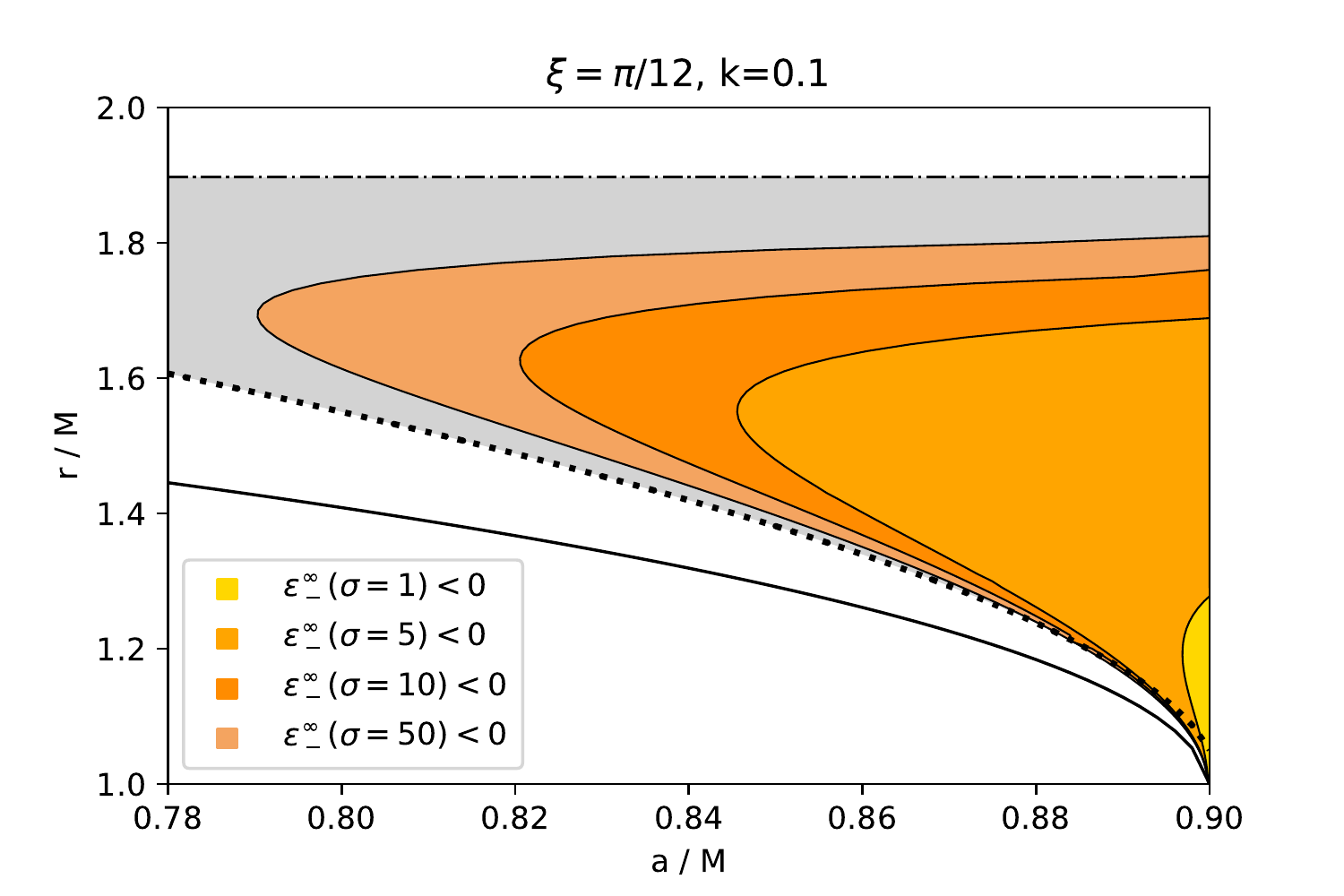}
\caption{The parameter space $(r-a)$ with $\xi=\pi/12$ and three different regular parameter $k=0$ (upper plot), 0.05(middle plot), 0.1(lower plot). The colored regions are $\epsilon_{-}^{\infty}<0$ with $\sigma_0=1, 5, 10, 50$. The gray area is the region where $\epsilon_{+}^{\infty}>0$. Black solid curves, black dotted curves, and black dot dashed curves are the radii of the outer event horizon, light ring, and outer ergosphere respectively.}
\label{ras}
\end{figure}

\begin{figure}[htbp]
  \includegraphics[scale=0.6]{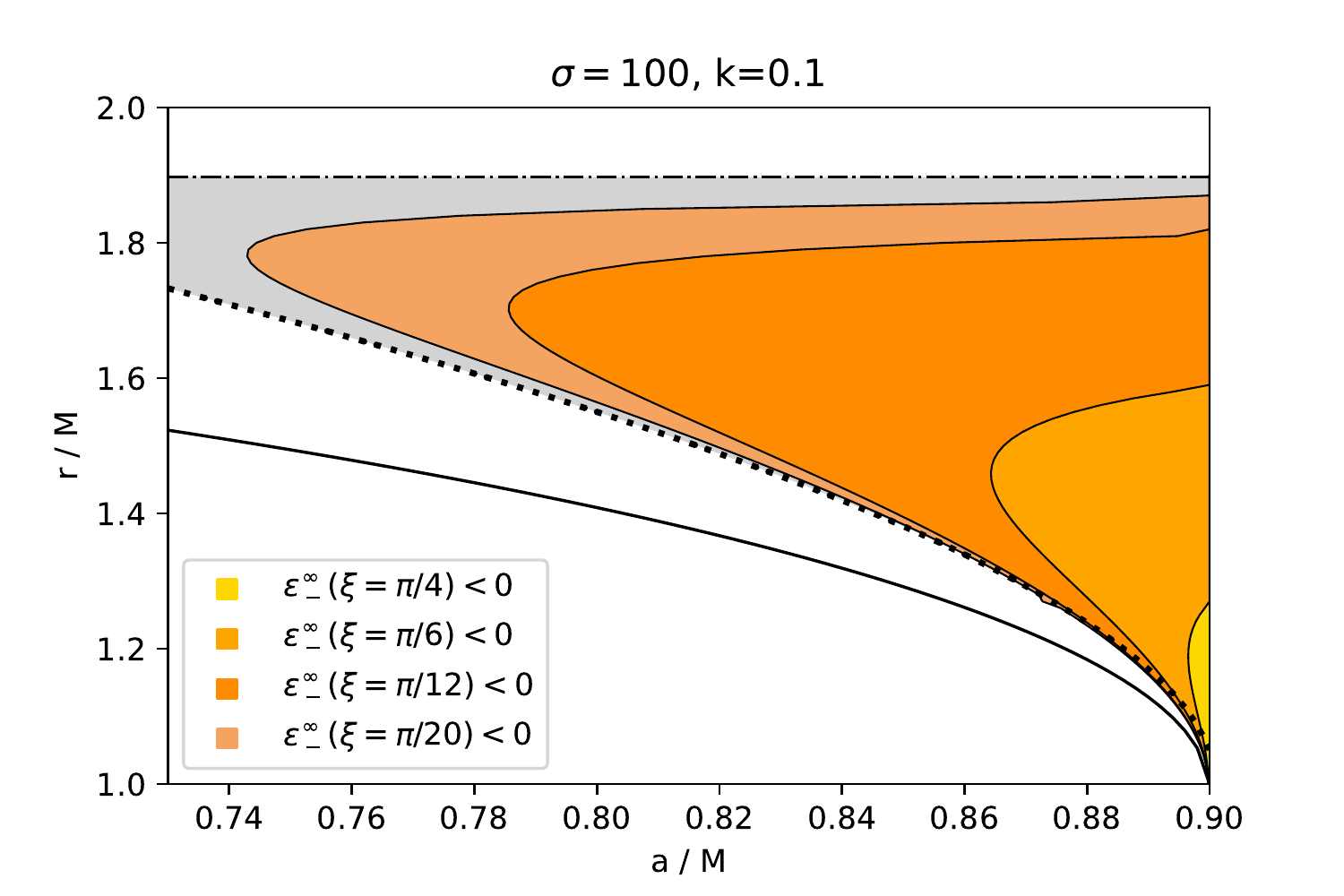}
\caption{The parameter space $(r-a)$ with $\sigma_0=100$, $k=0.1$ and different orientation angle $\xi$. The colored regions are $\epsilon_{-}^{\infty}<0$ with $\xi=\pi/20, \pi/12, \pi/6, \pi/4$ respectively. The gray area is the region where $\epsilon_{+}^{\infty}>0$. Black solid curves, black dotted curves, and black dot dashed curves are the radii of the outer event horizon, light ring, and outer ergosphere respectively.}
\label{rax}
\end{figure}

Let us now examine the parameter space of the dominant reconnection radial location $r$ and black hole spin $a$ that permit the realization of energy extraction from black hole via magnetic reconnection. The results are shown in two-dimensional $r - a$ parameter space plots, see Fig.\ref{ras} and Fig.\ref{rax}. In Fig.\ref{ras}, we have three plots, they correspond to different regular parameter $k=0,0.05,0.1$ respectively by taking $\xi=\pi/12$. The $k=0$ case is the Kerr black hole as reference. 
{The  critical black hole spin, under which the rotating black hole has two horizons,  decreases as  the regular parameter $k$ increases. We only show the region below those critical black hole spin. As the regular parameter $k$ increase, the radii of the outer event horizon and outer ergosphere decrease while that of light ring increase. }
The allowed region for $\epsilon_{+}^{\infty}>0$ (gray area) is widening in the $r$ dimension while shrinking in the $a$ dimension. The regions where $\epsilon_{-}^{\infty}<0$ have a great dependency on the plasma magnetization $\sigma_0$; the larger $\sigma_0$, the larger the region with  $\epsilon_{-}^{\infty}<0$. The increase in the regular parameter $k$ will make smaller the black hole spin $a$ available for $\epsilon_{-}^{\infty}<0$. Regarding to Fig.\ref{rax}, we aim to exam the effects of orientation angle $\xi$ on the $r - a$ parameter space by taking $\sigma_0=100$, $k=0.1$ . We can see that the smaller orientation angle $\xi$, the larger region with $\epsilon_{-}^{\infty}<0$. 

\section{Energy extraction power and efficiency}\label{sec5}
The power and efficiency of energy extraction via the Comisso-Asenjo mechanism are significant to the black hole evolution and its astrophysical phenomena. In this section, we will investigate the energy extraction power and efficiency of the rotating regular black hole (\ref{metric}). In \cite{mr11}, Comisso and Asenjo have proposed that these two quantities essentially depend on how quick the plasma with negative energy-at-infinity density are absorbed by the black hole per unit time. The power can be well estimated by \cite{mr11}
\begin{equation}
P_{etr}=-\epsilon_{-}^{\infty}w_0 A_{in} U_{in}
\end{equation}
where the reconnection inflow four-velocity $U_{in}=\mathcal{O}(10^{-1})$ and $\mathcal{O}(10^{-2})$ respectively refer to the collisionless and collisional regimes. $A_{in}$ is the cross-sectional area of the inflowing plasma, which can be estimated as $A_{in} \sim  r_E^2- r_{ph}^2 $ for rapid rotating regular black holes, with $r_E$ and $r_{ph}$ are the outer ergosphere and light ring  of the rotating regular black hole respectively.

\begin{figure}[htbp]
  \includegraphics[scale=0.6]{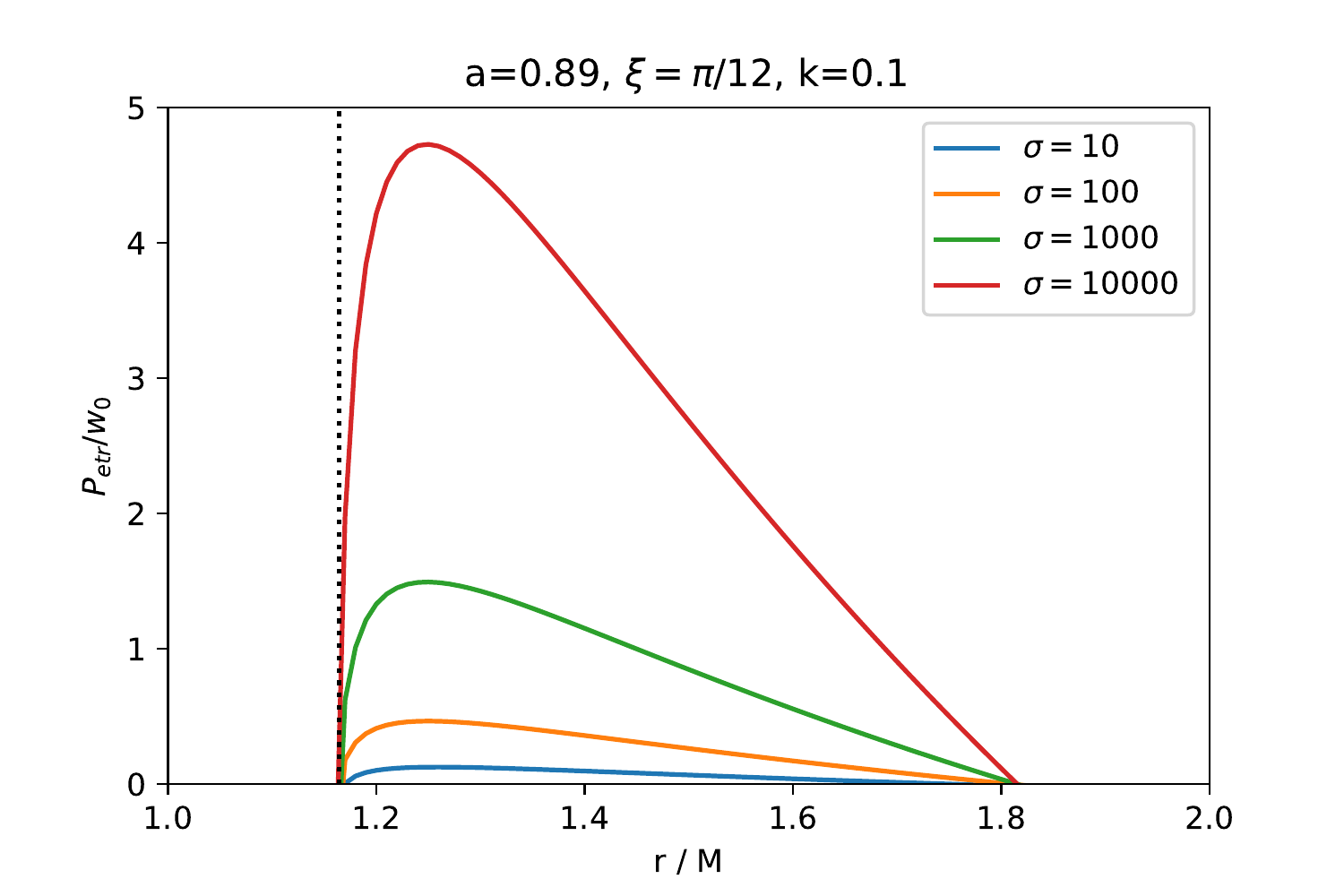}
\caption{$P_{etr}/w_0$ as a function of the dominant reconnection radial location $r$ with different plasma magnetization $\sigma_0=10,100,1000,10000$, by taking $a=0.89$, $\xi=\pi/12$, $k=0.1$ and $U_{in}=0.1$. The vertical dotted line indicates the limiting circular orbit, i.e., light ring $r_{ph}$. }
\label{power}
\end{figure}

\begin{figure}[htbp]
  \includegraphics[scale=0.6]{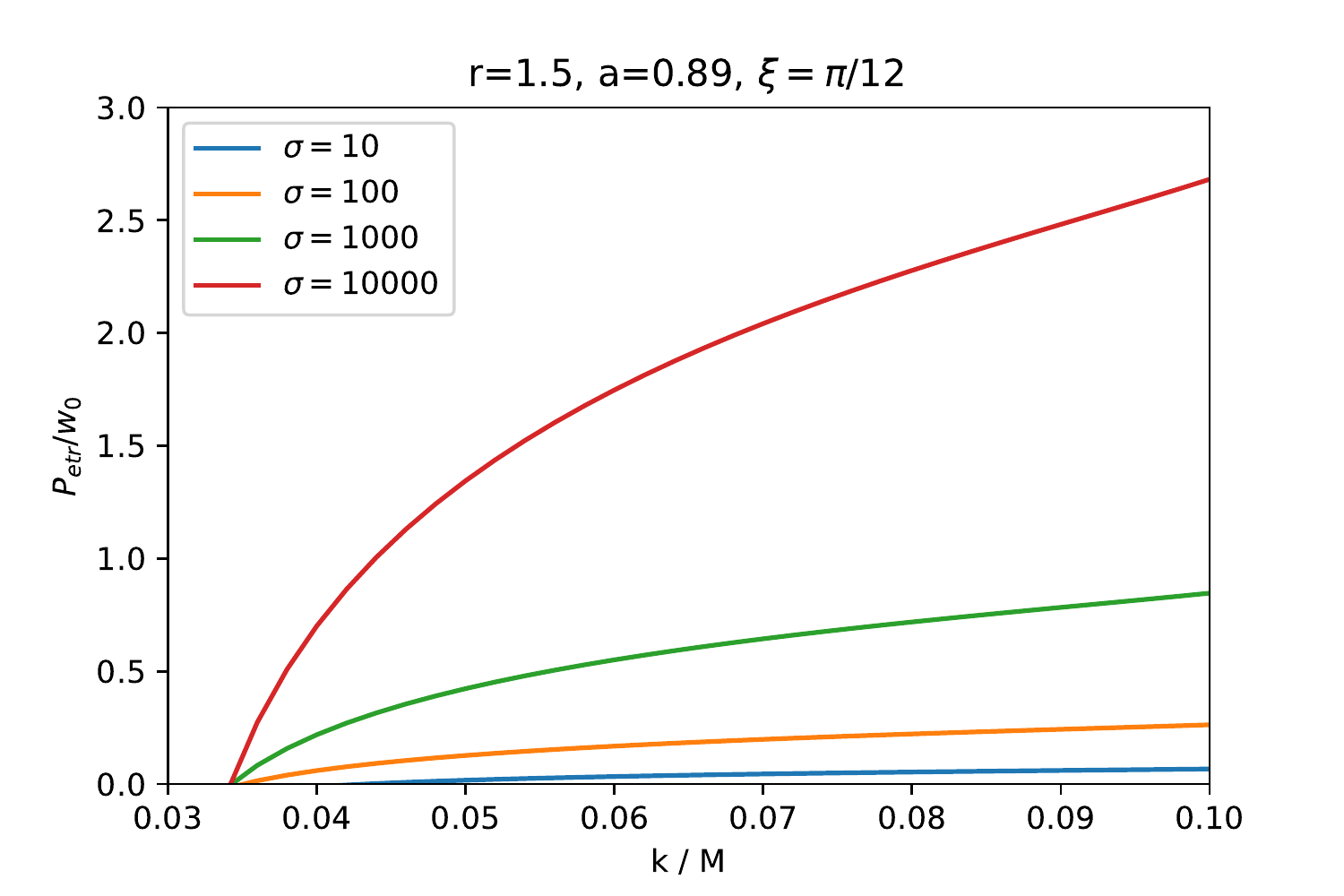}
\caption{ $P_{etr}/w_0$ as a function of the regular parameter $k$ with different plasma magnetization $\sigma_0=10,100,1000,10000$, by taking $r=1.5$, $a=0.89$, $\xi=\pi/12$ and $U_{in}=0.1$.}
\label{power-k}
\end{figure}

In Fig.\ref{power}, we demonstrate the ratio $P_{etr}/w_0$ as a function of the dominant reconnection radial location $r$ for a rapidly spinning black hole (\ref{metric}) in the collisionless regime $U_{in}=0.1$, with different plasma magnetization $\sigma_0=10,100,1000,10000$ by taking $a=0.89$, $\xi=\pi/12$  and $k=0.1$. As the plasma magnetization $\sigma_0$ increases, the power extracted from the black hole rises monotonically. The power peaks at those positions near the limiting circular orbit or light ring and then gradually declines. 

We also show the ratio $P_{etr}/w_0$ as a function of the regular parameter $k$ in Fig.\ref{power-k}. with different plasma magnetization $\sigma_0=10,100,1000,10000$ by taking $r=1.5$, $a=0.89$, $\xi=\pi/12$ and $U_{in}=0.1$. The power increases monotonically for the increasing  values of the regular parameter $k$. In addition, along with the plasma magnetization $\sigma_0$, the power rises monotonically as well.

Next, we evaluate the efficiency of energy extraction. It is convenient to define the efficiency as \cite{mr11}
\begin{equation}
\eta=\frac{\epsilon_{+}^{\infty}}{\epsilon_{+}^{\infty}+\epsilon_{-}^{\infty}}.
\end{equation}
If $\eta >1$, then the energy will be extracted from the rotating regular black hole.

\begin{figure}[t]
\includegraphics[scale=0.6]{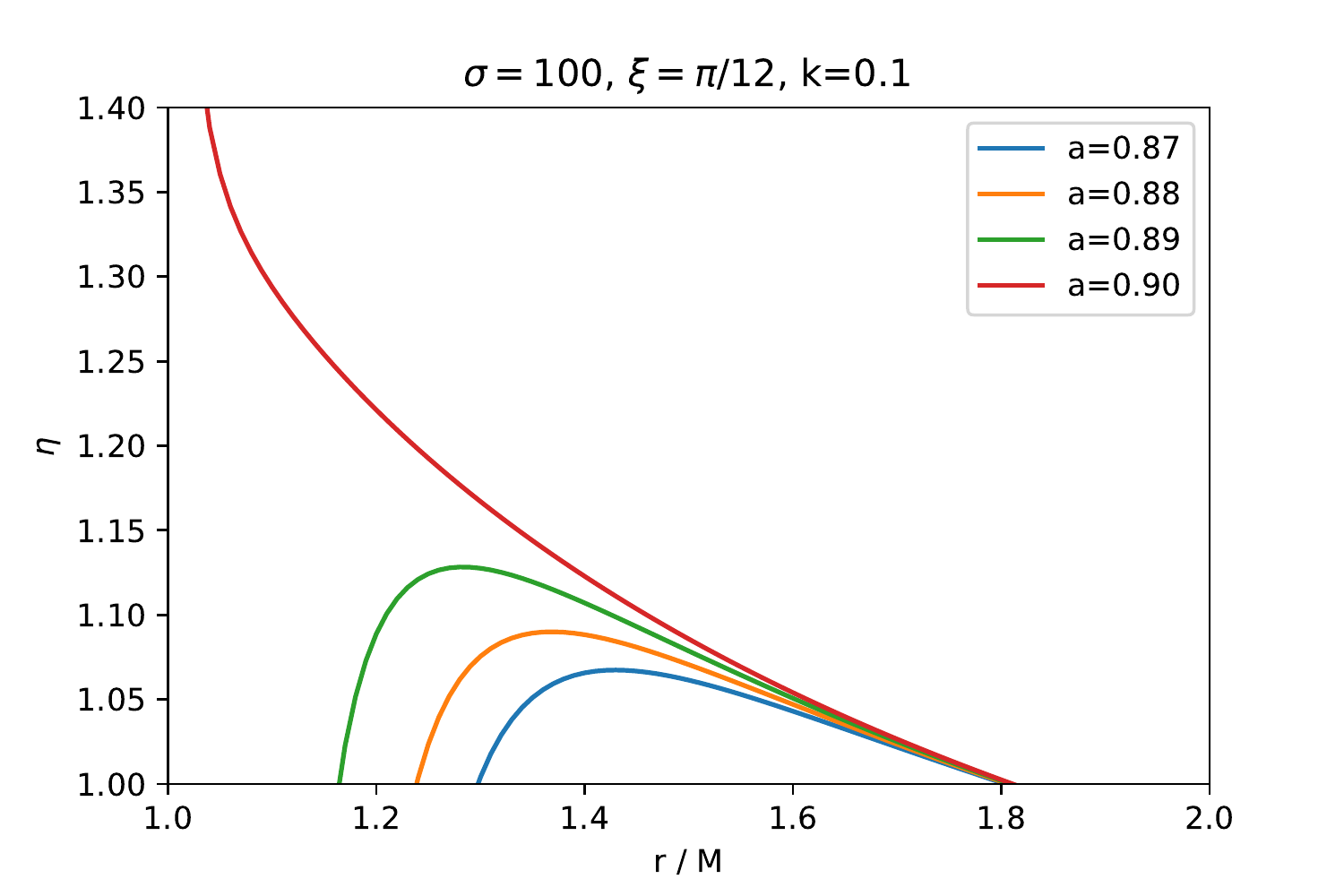}
\caption{Efficiency $\eta$ of the magnetic reconnection process as a function of the dominant reconnection radial location $r$ with different black hole spin $a=0.87,0.88,0.89,0.90$, taking $\sigma=100$, $\xi=\pi/12$, $k=0.1$. Note that the $a=0.90$ case is the extreme black hole for $k=0.10$.}
\label{eta}
\end{figure}

In Fig.\ref{eta}, we show the efficiency $\eta$ as a function of the dominant reconnection radial location $r$ with different black hole spin $a=0.87,0.88,0.89,0.90$, taking $\sigma=100$, $\xi=\pi/12$, $k=0.1$. Note that when $a=0.90$, the black hole is an extreme one for $k=0.10$. As reference, we also plot the efficiency for the extreme black hole. Regarding to the non-extremal case, we can see from Fig.\ref{eta} that the efficiency significantly increases with location $r$ that are closer to the outer event horizon and decreases below unity when it is close to the ergosphere. Thus, there are peaks for efficiency and the peaks go to large value and shift to small location $r$ when black hole spin $a$ increases.

\begin{figure}[htbp]
  \includegraphics[scale=0.6]{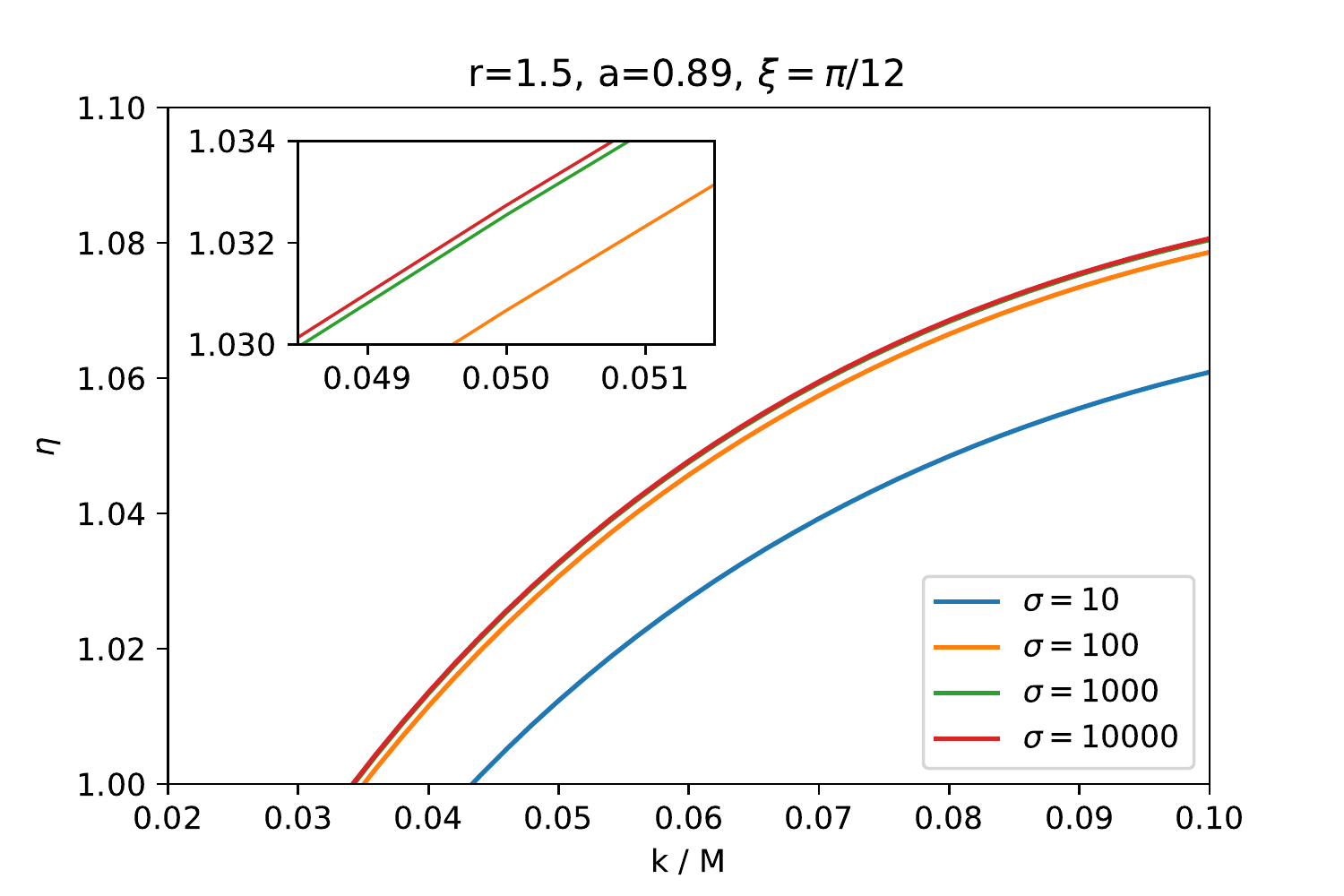}
\caption{Efficiency $\eta$ of the magnetic reconnection process as a function of the regular parameter $k$ with different plasma magnetization $\sigma_0=10,100,1000,1000$, taking $r=1.5$, $a=0.89$ $\xi=\pi/12$.}
\label{eta-k}
\end{figure}

In order to study the role of the regular parameter $k$ on the efficiency, we plot the efficiency as a function of the regular parameter $k$ with different plasma magnetization $\sigma_0=10,100,1000,1000$ in Fig.\ref{eta-k}, taking $r=1.5$, $a=0.89$ $\xi=\pi/12$. We can see that the efficiency grows monotonically along with the increasing of regular parameter $k$. The effects of plasma magnetization $\sigma_0$ on the efficiency is dropping exponetially as $\sigma_0$ grows and almost the same for sufficient large values.

Last but not least, we compare the powers of the energy extractions via Comisso-Asenjo and Blandford-Znajek mechanism, in the latter of which the black hole rotation energy is extracted through a magnetic field that threads the event horizon. Regarding to the Blandford-Znajek mechanism, in the so-called maximum efficiency conditions \cite{bz1,bz2,bz3}, the power of energy extraction is given by \cite{bz4,bz5,bz6}
\begin{equation}\label{bzz}
P_{BZ} =\frac{\kappa}{16\pi}\Phi_{H}^2(\Omega_{H}^2+C_1 \Omega_{H}^4+C_2\Omega_{H}^6+\mathcal{O}(\Omega_{H}^8))
\end{equation}
where $\kappa$ is a numerical constant related to the magnetic field configuration and  $C_1,C_2$ are numerical coefficients. The magnetic flux crossing the black hole event horizon is given by $\Phi_{H}=\frac{1}{2} \int_{\theta}\int_{\phi}\left | B_r \right | \sqrt{-g} d\theta d\phi=2\pi(r_+^2+a^2)B_0 sin(\xi)$, where $r_+$ is the event horizon. The angular velocity at the event horizon is $\Omega_{H}=2 a r_+e^{-k/r_+}/{\left(r_+^{2}+a^{2}\right)^{2}}$. We assume that the difference in the spacetime metric will only affect the Blandford-Znajek process through geometry quantities, since the basic magnetic field configurations are the same to the Kerr case. Thus, we have only modified the geometry quantities in the Blandford-Znajek power of Kerr black hole and obtain Eq.(\ref{bzz}). Then, the power ratio between these two mechanism is 
\begin{equation}\label{pr}
\frac{P_{etr}}{P_{BZ}}\sim \frac{-4\epsilon_{-}^{\infty} A_{in} U_{in}}{\pi \kappa \sigma_0(r_+^2+a^2)^2 sin^2(\xi) (\Omega_{H}^2+C_1 \Omega_{H}^4+C_2\Omega_{H}^6)}.
\end{equation}

\begin{figure}[htbp]
  \includegraphics[scale=0.6]{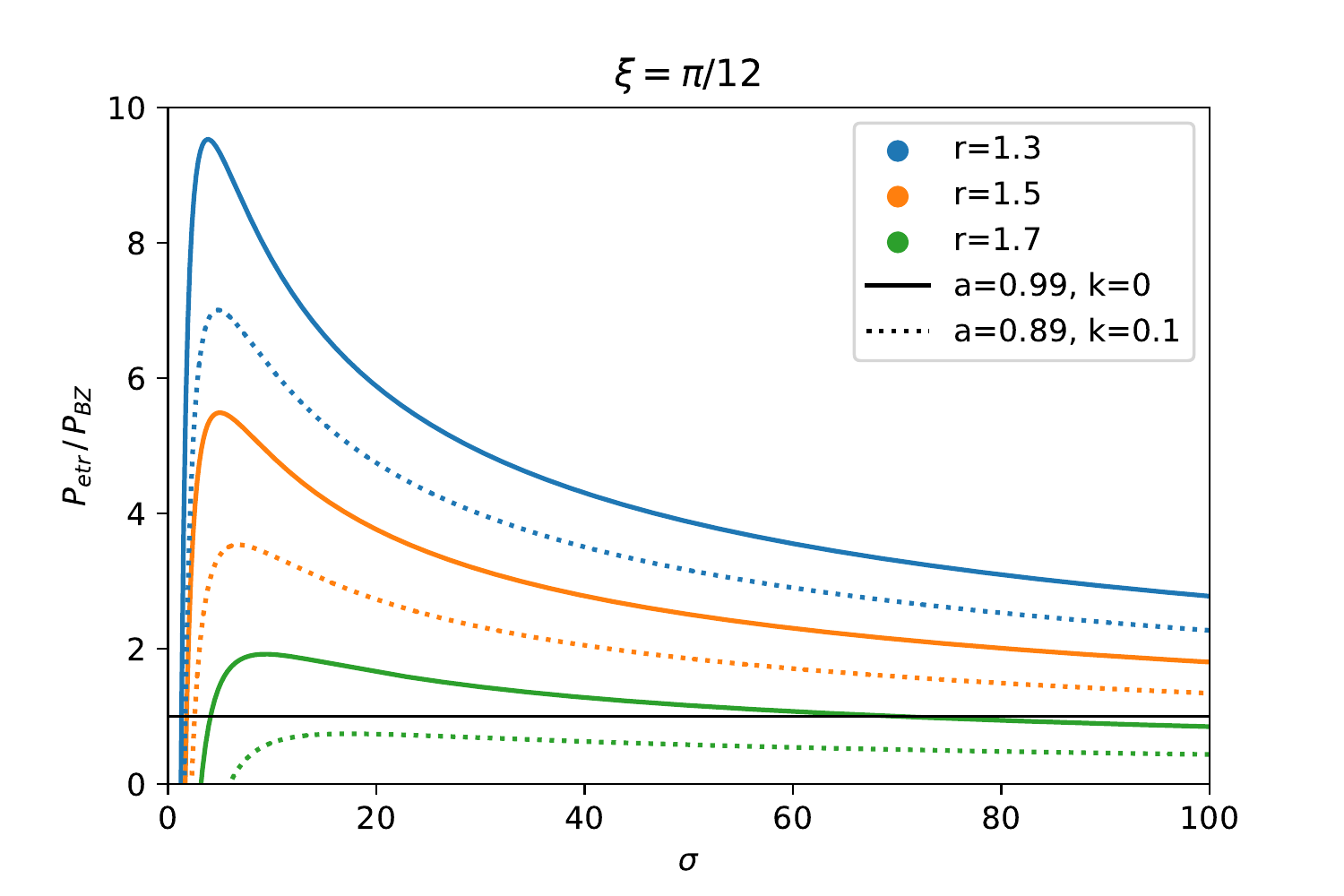}
\caption{Power ratio $P_{etr}/P_{BZ}$ as a function of the plasma magnetization $\sigma_0$ with different dominant reconnection radial location $r=1.3,1.5, 1.7$ and [$a=0.89, k=0.1$](dotted curve), [$a=0.99, k=0$] (solid curve), by taking the orientation angle $\xi=\pi/12$. The coefficients are taken as $\kappa=0.05$, $C_1=1.38$, $C_2=-9.2$ respectively. The black solid line is $P_{etr}/P_{BZ}=1$ as reference.}
\label{ratio}
\end{figure}

By taking $\xi=\pi/12$, $\kappa=0.05$, $C_1=1.38$, $C_2=-9.2$ respectively  \cite{mr11}, we show the power ratio in Fig.\ref{ratio} as a function the plasma magnetization $\sigma_0$ with different dominant reconnection radial location $r=1.3,1.5, 1.7$, the near-extremal black hole spin $a=0.89,0.99$ corresponding to regular parameter $k=0.1,0$ respectively. We can see that all the power ratios rise sharply along with the plasma magnetization that is closer to the critical value, and then after their maximum value are attained, the power ratios drop along with the plasma magnetization. The reason why the power ratios Eq.(\ref{pr}) decrease as the increase in the plasma magnetization is the different dependency on the plasma magnetization. Since when $\sigma_0 \to \infty$, we have $P_{etr}\sim \sigma_0^{1/2}$ and $P_{BZ}\sim \sigma_0$ , thus $P_{etr}/P_{BZ}\sim 1/\sigma_0^{1/2}$, which decrease as the increase in plasma magnetization. On the other hand, when $\sigma_0 \sim 1$, the force-free electrodynamics approximation \cite{bz4,bz5,bz6} of Blandford-Znajek power becomes invalid. The power ratios can only be seen as an effective comparison. We also found that the smaller the dominant reconnection radial location $r$, the larger the power ratios.  In addition, the power ratio of Kerr black hole ($k=0$) is greater than the regular black hole ($k=0.1$), which shows that, comparing with the Kerr black hole, the Comisso-Asenjo mechanism is less effective at extracting energy from a regular black hole. Nevertheless, in a very broad parameter range of the plasma magnetization
$\sigma_0$, the power ratio is greater than 1, which means that the Comisso-Asenj mechanism is a very promising and important energy extraction mechanism from rotating regular black holes.

\section{conclusion}\label{sec6}
Rotating regular black hole (\ref{metric}) is a very promising solution to the singularity problem in general relativity. In order to verify this proposal, the phenomenological study of this spacetime is essential. In this work, we investigated the energy extraction from a rotating regular black hole (\ref{metric}) caused by Comisso-Asenj magnetic reconnection process within the ergosphere. 

We first present the Comisso-Asenjo formulas of the plasma energy at infinity (\ref{ep}) associated with the accelerated and decelerated part of the same reconnection process as well as the conditions (\ref{cd}) for extracting energy from the rotating regular black hole. 

Then we specifically studied how the plasma magnetization, orientation angle and especially the regular parameter will affect the behaviour of the plasma energy-at-infinity per enthalpy $\epsilon_{+}^{\infty}$ and $\epsilon_{-}^{\infty}$, by taking black hole spin and dominant reconnection radial location as certain values. It turns out that the small value of orientation angle and large values of plasma magnetization and regular parameter will be beneficial for the energy extraction via magnetic reconnection from the rotating regular black hole. What's more, in order to explore the effects of the other parameters settled down, we demonstrated the two dimensional $(r-a)$ parameter space that satisfy the conditions (\ref{cd}). We also show the effects of different regular parameter, plasma magnetization and orientation angle in the same figures such that we can fully explore the parameter spaces. 

In the subsequent section, we also studied how these critical parameters will affect the power and efficiency of the energy extraction via the magnetic reconnection. 
The power and efficiency all grow with the increasing of regular parameter. It is interesting that the plasma magnetization almost has no effect to the efficiency when it becomes large enough.

At last, we studied the power ratio of the energy extraction via the Comisso-Asenjo mechanism to the famous Blandford-Znajek mechanism. 
In a sufficient large parameter range, the energy extraction via Comisso-Asenjo process is more powerful than the Blandford-Znajek mechanism for extracting energy from the rotating regular black hole. In addition, the Kerr case $(k=0)$ is more efficient than the regular black hole case $(k=0.1)$. 

\section*{Acknowledgments}
Zhen Li would like to thank the DARK cosmology centre at Niels Bohr Institute for supporting this research. Zhen Li is also financially supported by the China Scholarship Council. Faqiang Yuan is supported by the National Natural Science Foundation of China (Grant Nos.11875006, and 11961131013). We thank Filippo Camilloni for helpful discussions.  We also would like to thank Luca Comisso for his suggestions and comments which has significantly improved this paper.
\\
\\


\begin{thebibliography}{0}%
\makeatletter
\providecommand \@ifxundefined [1]{%
 \@ifx{#1\undefined}
}%
\providecommand \@ifnum [1]{%
 \ifnum #1\expandafter \@firstoftwo
 \else \expandafter \@secondoftwo
 \fi
}%
\providecommand \@ifx [1]{%
 \ifx #1\expandafter \@firstoftwo
 \else \expandafter \@secondoftwo
 \fi
}%
\providecommand \natexlab [1]{#1}%
\providecommand \enquote  [1]{``#1''}%
\providecommand \bibnamefont  [1]{#1}%
\providecommand \bibfnamefont [1]{#1}%
\providecommand \citenamefont [1]{#1}%
\providecommand \href@noop [0]{\@secondoftwo}%
\providecommand \href [0]{\begingroup \@sanitize@url \@href}%
\providecommand \@href[1]{\@@startlink{#1}\@@href}%
\providecommand \@@href[1]{\endgroup#1\@@endlink}%
\providecommand \@sanitize@url [0]{\catcode `\\12\catcode `\$12\catcode
  `\&12\catcode `\#12\catcode `\^12\catcode `\_12\catcode `\%12\relax}%
\providecommand \@@startlink[1]{}%
\providecommand \@@endlink[0]{}%
\providecommand \url  [0]{\begingroup\@sanitize@url \@url }%
\providecommand \@url [1]{\endgroup\@href {#1}{\urlprefix }}%
\providecommand \urlprefix  [0]{URL }%
\providecommand \Eprint [0]{\href }%
\providecommand \doibase [0]{https://doi.org/}%
\providecommand \selectlanguage [0]{\@gobble}%
\providecommand \bibinfo  [0]{\@secondoftwo}%
\providecommand \bibfield  [0]{\@secondoftwo}%
\providecommand \translation [1]{[#1]}%
\providecommand \BibitemOpen [0]{}%
\providecommand \bibitemStop [0]{}%
\providecommand \bibitemNoStop [0]{.\EOS\space}%
\providecommand \EOS [0]{\spacefactor3000\relax}%
\providecommand \BibitemShut  [1]{\csname bibitem#1\endcsname}%
\let\auto@bib@innerbib\@empty
\end{thebibliography}%


\begin{thebibliography}{100}
\bibitem{gw1}
B. P. Abbott et al. (LIGO Scientific, Virgo), Phys. Rev. Lett. {\bf 116}, 061102 (2016).
\bibitem{gw2}
B. P. Abbott et al. (LIGO Scientific, Virgo), Phys. Rev. D. {\bf 100}, 104036 (2019).
\bibitem{gw3}
B. P. Abbott et al. (LIGO Scientific, Virgo), Phys. Rev. Lett. {\bf 119}, 161101 (2017).
\bibitem{shadow1}
K. Akiyama et al. (Event Horizon Telescope), Astrophys. J. Lett. {\bf 875}, L1 (2019).
\bibitem{shadow2}
K. Akiyama et al. (Event Horizon Telescope), Astrophys. J. Lett. {\bf 930}, L15 (2022).
\bibitem{cha2}
R. Penrose, Phys. Rev. Lett. {\bf 14}, 57 (1965).
\bibitem{cha3}
S. W. Hawking and R. Penrose, Proc. Roy. Soc. Lond. A  {\bf 314}, 529 (1970).
\bibitem{re}
J. M. Bardeen, in {\it Conference Proceedings of GR5}, p. 174. Tbilisi University Press, 1968..
\bibitem{re2}
E. Ay\'on-Beato and A. Garcia, Phys. Lett. B {\bf493}, 149 (2000).
\bibitem{Hay06}  
S. Hayward, Phys. Rev. Lett. {\bf96}, 031103 (2006).
%
\bibitem{LYGM23} 
C. Lan, H. Yang, Y. Guo, and Y.-G. Miao,  arXiv:2303.11696v2.
\bibitem{nre} 
X. Li, Y. Ling, and Y.-G. Shen, Int. J. Mod. Phys. D {\bf22},  1342016 (2013).
\bibitem{nre1}
H. Culetu, arXiv:1305.5964v6; 1508.07570v6.
\bibitem{nre2}
H. Culetu. Int. J. Theor. Phys. {\bf 54}, 2855 (2015).
\bibitem{Cul17} H. Culetu, J. Phys.: Conf. Ser. {\bf845}, 012006 (2017).
\bibitem{Cul22} H. Culetu,  Int. J. Mod. Phys. D {\bf31}, 2250124 (2022).
\bibitem{Cul23} H. Culetu,   Phys. Lett. B {\bf839}, 127775 (2023).
\bibitem{nre5} S. G. Ghosh, Eur. Phys. J. C. {\bf 75}, 532 (2015).
\bibitem{nre4} A. Simpson and M. Visser, Universe. {\bf 6}, 8 (2020).
\bibitem{nre6} A. Simpson and M. Visser. JCAP03(2022)011.
\bibitem{nre7} A. Simpson and M. Visser. Phys. Rev. D {\bf 105}, 064065 (2022).
\bibitem{re4}C. Bambi and L. Modesto, Phys. Lett. B. {\bf 721}, 329 (2013).
\bibitem{fw1}
M. Amir and S. G. Ghosh, Phys. Rev. D {\bf94}, 024054 (2016).
\bibitem{fw2}
R.~Kumar and S.~G.~Ghosh, Class. Quant. Grav. {\bf38}, 8 (2021).
\bibitem{fw3}
R.~Kumar, A.~Kumar and S.~G.~Ghosh,
Astrophys. J. {\bf 896}, 89 (2020).
\bibitem{SAAA22} 
F. Sarikulov, F. Atamurotov, A. Abdujabbarov, and B. Ahmedov, Eur. Phys. J. C {\bf82}, 771 (2022).
\bibitem{fw4}
Z. Li, Phys. Rev. D {\bf 107}, 044013 (2023).
\bibitem{mr1}
 W. Daughton, V. Roytershteyn, B. Albright, H. Karimabadi, L. Yin, and K. J. Bowers, Phys. Rev. Lett. {\bf 103}, 065004 (2009).
\bibitem{mr2}
A. Bhattacharjee, Y.-M. Huang, H. Yang, and B. Rogers, Phys. Plasmas {\bf 16}, 112102 (2009).
\bibitem{mr3}
K. Parfrey, A. Philippov and B. Cerutti, Phys. Rev. Lett. {\bf 122}, 035101 (2019).
\bibitem{mr4}
S. S. Komissarov, Mon. Not. Roy. Astron. Soc. {\bf 359}, 801 (2005).
\bibitem{mr5}
B. Ripperda, F. Bacchini and A. Philippov, Astrophys. J. {\bf 900}, 100 (2020).
\bibitem{mr6}
A. Bransgrove, B. Ripperda and A. Philippov, Phys. Rev. Lett. {\bf 127}, 055101 (2021).
\bibitem{mr11}
L. Comisso and F. A. Asenjo, 
Phys. Rev. D {\bf 103}, 023014 (2021).
\bibitem{mr7}
L. Comisso, M. Lingam, Y.-M. Huang, and A. Bhattacharjee,  Phys. Plasmas {\bf 23}, 100702 (2016)
\bibitem{mr8}
D. A. Uzdensky and N. F. Loureiro, Phys. Rev. Lett. {\bf 116}, 105003 (2016)
\bibitem{mr9}
L. Comisso, M. Lingam, Y.-M. Huang, and A. Bhattacharjee, Astrophys. J. {\bf 850}, 142 (2017).
\bibitem{mr10}
S. Koide and K. Arai, Aprophys. J. {\bf 682}, 1124 (2008).
\bibitem{p1}
R. Penrose and R. M. Floyd. Nature 
{\bf 229}, 177 (1971).
\bibitem{p2}
T. Piran, J. Shaham, and J. Katz. Astrophys. J. Lett. {\bf 196}, L107 (1975).
\bibitem{bz4}
R. D. Blandford and R. L. Znajek, Mon. Not. R. Astron. Soc. {\bf179}, 433 (1977).
\bibitem{mrr1}
S. W. Wei, H. M. Wang, Y. P. Zhang, and Y. X. Liu, JCAP {\bf 04}, 050 (2022).
\bibitem{mrr2}
W. Liu, Astrophys. J. {\bf 925}, 149 (2022).
\bibitem{mrr3}
M. Khodadi, Phys. Rev. D {\bf 105}, 023025 (2022).
\bibitem{mrr4}
A. Carleo, G. Lambiase, and L. Mastrototaro, Eur. Phys. J. C {\bf 82}, 776 (2022).
\bibitem{mrr5}
C.~H.~Wang, C.~Q.~Pang and S.~W.~Wei, Phys. Rev. D {\bf 106}, 124050 (2022).
\bibitem{zamo}
J. M. Bardeen, W. H. Press, and S. A. Teukolsky, Astrophys. J. {\bf 178}, 347 (1972).
\bibitem{AC17} 
F. A. Asenjo and L. Comisso, Phys. Rev. Lett. {\bf118}, 055101 (2017).
\bibitem{bz1}
D. Macdonald and K. S. Thorne, Mon. Not. R. Astron. Soc. {\bf198}, 345 (1982).
\bibitem{bz2}
K. S. Thorne, R. H. Price, and D. A. MacDonald, Black Holes: The Membrane Paradigm (Yale University,
New Haven, CT, 1986).
\bibitem{bz3}
S. S. Komissarov, Mon. Not. R. Astron. Soc. {\bf 326}, L41 (2001).
\bibitem{bz5}
A. Tchekhovskoy, R. Narayan, and J. C. McKinney, As-trophys. J. {\bf711}, 50 (2010).
\bibitem{bz6}
Filippo Camilloni et al. JCAP {\bf 07}, 032 (2022).
\end{thebibliography}
\end{document}